\documentclass[submission,copyright,creativecommons]{eptcs}
\providecommand{\event}{QPL  2016} 
\usepackage{breakurl}             
\usepackage{underscore}           

\usepackage{amssymb, amsmath} 
\usepackage[arrow,matrix,curve,cmtip,ps]{xy}
\usepackage{xcolor}
\newcommand{\red}{}

\usepackage{tikz}
\usepackage{graphicx}
\tikzset{ lines/.style=very thin}

\newcommand{\tempout}[1]{{}}

\renewcommand*{\dot}{\raisebox{-0.88ex}{\scalebox{2.5}{$\cdot$}}}

\newcommand{\E}{{\mathbf E}}

\renewcommand{\L}{{\mathcal L}}
\newcommand{\M}{\boldsymbol{\mathcal M}}

\newcommand{\V}{{\mathbf V}}

\newcommand{\ring}{{\mathbb K}}
\newcommand{\R}{{\mathbb R}}
\newcommand{\C}{{\mathbb C}}

\newcommand{\Q}{{\mathbb H}}

\renewcommand{\O}{{\mathbb O}}

\renewcommand{\H}{\boldsymbol{\mathcal H}}

\renewcommand{\1}{{\mathbf 1}}

\newcommand{\Tr}{\mbox{Tr}}

\renewcommand{\hat}{\widehat}

\renewcommand{\bar}{\overline}

\newcommand{\Cat}{{\mathcal C}}

\newcommand{\FdHilb}{\mathbf{FdHilb}}

\newcounter{thaler}
\newenvironment{mlist}{\begin{list}{\arabic{thaler}}%
{\usecounter{thaler}
\setlength{\rightmargin}{\leftmargin}
\topsep=0pt
\itemsep=0pt
\parskip=0pt
\parsep=0pt
}}{\end{list}}
\title{A Royal Road to Quantum Theory\\  (or Thereabouts)  \\
Extended Abstract} 
\author{Alexander Wilce
\institute{Susquehanna University\\Selinsgrove, PA}
\email{wilce@susqu.edu}
}
\def\titlerunning{A Royal Road to QM}
\def\authorrunning{A. Wilce}
\begin{document}
\maketitle

\begin{abstract} 
A representation of finite-dimensional probabilistic models in terms of formally real Jordan algebras is obtained, 
in a strikingly easy way, from simple assumptions. This provides a framework in which real, complex and quaternionic 
quantum mechanics can be treated on an equal footing, and allows some (but not too much) room for other alternatives. 
This is based on earlier work (arXiv:1206:2897), but the development here is further simplified, and also extended in several ways. I also discuss the possibilities for 
organizing probabilistic models, subject to the assumptions discussed here, into symmetric monoidal categories, 
showing that such a category will automatically have a dagger-compact structure. (Recent joint work with Howard Barnum and Matthew Graydon (arXiv:1507.06278) exhibits several categories of this 
kind.)  
\end{abstract}

\section{Introduction and Overview} 

Several recent papers, notably \cite{CDP, Dakic-Brukner, Masanes-Mueller}, have derived the standard formulation of finite-dimensional quantum mechanics (QM)  
from various packages of axioms governing the information-carrying and information-processing capacity of finite-dimensional probabilistic systems.  In this paper, I derive somewhat less, but do so in what I think is a very attractive and simple way.  Specifically, I characterize {\em formally real Jordan algebras} as probabilistic models, 
in terms of a few assumptions having straightforward probabilistic interpretations. 
This allows some --- but not too much --- latitude to go beyond standard finite-dimensional  complex QM. (All simple formally real Jordan algebras are self-adjoint parts either of matrix algebras $M_n(\ring)$, where $\ring$ is either $\R, \C$ or $\Q$, or, if $n = 3$, $\O$ (the Octonions), or of Clifford algebras \cite{JNW}. The first three cases correspond to finite-dimensional real, complex and quaternionic quantum-mechanical systems.) 
Moreover, this approach is significantly simpler mathematically than any of those cited above. 
Once the various definitions are in place, the proofs of the main theorems are all quite easy, at least if one is allowed to invoke one classical mathematical result. An ordered vector space $\E$, with positive cone $\E_+$, is {\em self-dual} iff there exists an inner product on $\E$ such that $a \in \E_+$ iff $\langle a, b \rangle \geq 0$ for all $b \in \E_+$. Call $\E$ {\em homogeneous} iff the group of order automorphisms --- positive linear automorphisms with positive inverses --- on $\E$ acts transitively on the {\em interior} of $\E_+$. Any formally real Jordan algebra, ordered by its cone of squares $\E_+ := \{ a \dot a | a \in \E\}$, is homogeneous and self-dual. For an accessible proof of the following, see \cite{FK}. \\

\noindent{\bf Theorem [Koecher 1958; Vinberg 1961]:} {\em Let $\E$ be a finite-dimensional ordered vector 
space with a distinguished order-unit $u$. If $\E$ is homogeneous and self-dual, then there exists a unique product $\dot$ on $\E$ such that $(\E,\dot)$ is a formally real Jordan algebra with Jordan unit $u$, and $\E_+$ is the cone of squares.}\\ 

It is standard to represent an abstract probabilistic physical system in terms of an order unit space $(\E,u)$ in such a way that elements of $\E_+$ dominated by $u$ represent {\em effects} (essentially: measurement outcomes).   
Thus, if we can find a conceptually compelling way to motivate the homogeneity and self-duality of $\E$, we will have cleared a route to (the vicinity of) quantum theory. 

Much of what follows is drawn from the earlier papers \cite{Wilce09, Wilce11, Wilce12}, but the approach 
sketched here is organized somewhat differently, is (even) simpler, and goes somewhat further. 

\section{Probabilistic Models} 

A {\em test space} is a collection $\M$ of non-empty sets, regarded as 
the outcome-sets of various experiments, measurements, etc. We refer to a set $E \in \M$ as a test. 
Let $X = \bigcup \M$ be the {\em outcome-space} of $\M$. A {\em probability weight} on $\M$ is a 
mapping $\alpha : X \rightarrow [0,1]$ summing to $1$ on every test. We say that $\alpha$ is 
{\em non-singular} iff $\alpha(x) > 0$ for all $x \in X(A)$.\footnote{Material in this section is standard. See \cite{BarnumWilceFoils} for a more detailed account, and for further references. A good general reference 
for ordered vector spaces is \cite{AT}}\\

\noindent{\bf Definition:} A {\em probabilistic model} is a pair $A = (\M(A),\Omega(A))$ where $\M(A)$ is a test space and 
$\Omega(A)$ is a specified convex set of probability weights on $\M(A)$, called the {\em states} of the model. 
 \\

It is harmless to assume that, for every outcome $x \in X(A)$, there exists at least one state $\alpha \in \Omega(A)$ 
with $\alpha(x) > 0$. 
The span of $\Omega(A)$ in $\R^{X(A)}$, ordered by the cone $\V_+(A)$ of non-negative multiples of 
states, is denoted $\V(A)$. It will be useful below to note that the interior of the cone $\V(A)_+$ 
consists of positive multiples of non-singular states.
There is a unique positive functional $u_A \in \V(A)^{\ast}$ given by 
$u(\alpha) = 1$ for all $\alpha \in \Omega$. An {\em effect} on $\V(A)$ is positive linear functional $a \in \V^{\ast}$ with $0 \leq a \leq u$. For example, each outcome $x \in X$ defines an effect $\hat{x} \in \V(A)^{\ast}$ 
by evaluation, i.e., $\hat{x}(\alpha) = \alpha(x)$ for all $\alpha \in \V(A)$. Effects can be taken 
to represent outcomes of {\em mathematically} possible measurements, but I make no assumption about which effects, other than those of the form $\hat{x}$, are physically 
accessible.  \\

\noindent{\bf Classical, Quantum and Jordan Models} If $E$ is a finite set, the corresponding {\em classical model} is $A(E) = (\{E\}, \Delta(E))$ where $\Delta(E)$ is the simplex of probability weights on $E$. If $\H$ is a finite-dimensional complex Hilbert space, let $\M(\H)$ denote the set of orthonormal bases of $\H$: then $X = \bigcup \M(\H)$ is the unit sphere of $\H$, and any density operator $W$ on $\H$ defines a probability weight $\alpha_W$, given by $\alpha_W(x) = \langle W x, x \rangle$ for all $x \in X$. Letting 
$\Omega(\H)$ denote the set of states of this form, we obtain the {\em quantum model} $A(\H) = (\M(\H), \Omega(\H))$ associated with $\H$.  The space $\V(A(\H))$ is isomorphic to the 
space $\L_h(\H)$ of hermitian operators on $\H$, ordered as usual. 

More generally, every formally real 
Jordan algebra $\E$ gives rise to a probabilistic model. Recall  
that a Jordan algebra is a real commutative algebra $(\E,\dot)$ with unit element $u$, 
$\dot$ satisfying the Jordan identity $a^2 \dot (a \dot b) = a \dot (a^2 \dot b)$. $\E$ is 
{\em formally real} if $\sum_i a_{i}^2 = 0$ implies $a_i = 0$  for all $i$. 
A minimal or {\em primitive} idempotent of $\E$ is an element $p \in \E$ with 
$p^2 = p$ and, for $q = q^2 < p$, $q = 0$. A {\em Jordan frame} is a maximal pairwise orthogonal set of primitive idempotents. Let $X(\E)$ be the set of primitive idempotents, $\M(\E)$, the set of Jordan frames, and 
$\Omega(\E)$, the set of probability weights of the form $\alpha(p) = \langle a, p \rangle$ where $a \in \E_+$ with 
$\langle a, u \rangle = 1$. This  defines the {\em Jordan model} $A(\E)$ associated with $\E$. Where $\E = \L_{h}(\H)$ for a finite-dimensional Hilbert space $\H$, 
this {\em almost} recovers the quantum model $A(\H)$, 
except that  
we replace unit vectors by their associated projection operators, thus conflating outcomes that differ 
by a phase. \\

\noindent{\bf Sharp models} A probabilistic model $A$ is {\em sharp} iff, for every outcome $x \in X(A)$, there exists a 
unique state $\alpha \in \Omega(A)$ with $\alpha(x) = 1$. Quantum models are evidently sharp; more generally, 
any Jordan model is sharp. 
If $A$ is sharp, then there is a sense in which each test $E \in \M(A)$ is {\em maximally informative:} if we know 
for certain which outcome will occur, then we know the system's state exactly, as there is only one state in which this 
outcome has probability  $1$. Conversely, sharpness can be understood as the requirement that all tests be maximally 
informative in this sense. \\

\noindent{\bf Processes} We may want to regard two systems, represented by models $A$ and $B$,
as the input to and output from some {\em process}, whether dynamical or purely information-theoretic, that 
has some probability to destroy the system or otherwise ``fail".  
Such a process can be represented mathematically by a positive linear mapping $T: \V(A) \rightarrow \V(B)$ 
 taking each normalized state $\alpha$ of $A$ to a possibly {\em sub-normalized} state $T(\alpha)$ of $B$, i.e., 
$T(\alpha) = p \beta$ where $\beta \in \Omega(B)$ and $p \in [0,1]$ is the probability for the process 
to fail, given input state $\alpha$. 
I do not suppose that every positive linear mapping of this sort represents a physically possible process.

Even if a process $T$ has a nonzero probability of failure, it may be possible to reverse its effect with nonzero 
probability: \\

\noindent{\bf Definition:} A process $T : A \rightarrow B$ is {\em probabilistically reversible}, 
or {\em p-reversible}, for short, iff there exists a process $S$ such that, for all $\alpha \in \Omega(A)$, $(S \circ T)(\alpha) = p \alpha$, where $p \in (0,1]$. \\

This means that there is a probability $1-p$ of the process $S \circ T$ failing, but a probability 
$p$ that it will leave the system in its initial state. (Note that here, $p$ is independent of $\alpha$, 
since $\alpha \mapsto p\alpha$ is linear.) If  $T$ preserves normalization, so that $T(\Omega(A)) \subseteq \Omega(B)$, $S$ will also preserve normalization, and will undo the result of $T$ with probability 1. In this case, we just say that $T$ is {\em reversible}. 

Every process $T : \V(A) \rightarrow \V(B)$ has a dual mapping $T^{\ast} : \V^{\ast}(B) \rightarrow \V^{\ast}(A)$, 
also positive, given by $T^{\ast}(b)(\alpha) = b(T(\alpha))$ for all $b \in \V^{\ast}(B)$ and $\alpha \in \V(A)$. That $T$ takes normalized states to subnormalized states is equivalent to the requirement that 
$T^{\ast}(u_B) \leq u_A$, that is, that $T^{\ast}$ map effects to effects.  \\

\noindent{\bf Bipartite States} If $A$ and $B$ are two models, a (non-signaling) {\em bipartite state} on $A$ and $B$ is a mapping $\omega : X(A) \times X(B) \rightarrow \R$
such that (i) $\sum_{(x,y) \in E \times F} \omega(x,y) = 1$ for all $E \in \M(A), F \in \M(B)$; 
(ii) the {\em marginals}  
\[\omega_1(x) = \sum_{y \in E} \omega(x,y) \ \mbox{and} \ \omega_2(y) = \sum_{x \in E} \omega(x,y)\]
are independent of $E \in \M(A)$ and $F \in \M(B)$, respectively;
(iii) the {\em conditional states} 
\[\omega_{1|y}(x) := \omega(x,y)/\omega_2(y) \ \mbox{and} \ \omega_{2|x} := \omega(x,y)/\omega_{1}(x)\]
belong to $\Omega(A)$ and $\Omega(B)$, respectively.
We then have the {\em law of total probability}: 
\[\omega_1(x) = \sum_{y \in F} \omega_2(y) \omega_{1|y}(x) \ \mbox{and} \ \omega_{2}(y) = \sum_{x \in E} \omega_1(x) \omega_{2|x}(y).\](Note this implies that 
$\omega_1$ and $\omega_2$ belong to $\Omega(A)$ and $\Omega(B)$.)\\

\noindent{\bf The space $\E(A)$} At this point, it is helpful to introduce another ordered vector space associated with a model $A$. 
Let $\E(A)_+ \subseteq \V(A)^{\ast}$ be the set of all linear combinations $\sum_{i} t_i \hat{x}_i$ 
with $x_i \in X(A)$ and $t_i \geq 0$ for all $i$. This is a convex generating cone for $\V(A)^{\ast}$, generally smaller than the dual cone $\V(A)^{\ast}$. Write $\E(A)$ for the space $\V(A)^{\ast}$, as ordered by this smaller cone. The main utility of $\E(A)$ is the following observation:\\

\noindent{\bf Lemma 0:} {\em If $\omega$ is a bipartite state on $A$ and $B$, there exists a unique positive linear 
mapping $\hat{\omega} : \E(A) \rightarrow \V(B)$ such that $\hat{\omega}(\hat{x})(y) = \omega(x,y)$ 
for all $x \in X(A)$ and $y \in X(B)$. }\\

The proof is straightforward (consider the map $X(B) \rightarrow \V(B)$ defined by $y \mapsto \omega( \ \cdot \, y)$; then dualize). 
Since $\hat{\omega}(\hat{x}) = \omega_1(x) \omega_{2|x}$, I call $\hat{\omega}$ the {\em conditioning map} 
associated with $\hat{\omega}$.

\section{Self-duality and homogeneity for quantum models}

I will say that a probabilistic model $A$ is {\em self-dual} if there exists an inner product on $\E(A)$ 
with respect to which $\E(A)$ is self-dual {\em and} $\E(A)_+ \simeq \V(A)_+$ (in the sense that $a \in \E(A)_+$ iff 
$\exists \alpha \in \V(A)_+$ with 
$\langle a, b \rangle = \alpha(b)$ for all $b \in \E(A)$).  If $\V(A)_+$ --- and hence, also $\E(A)_+$ --- is 
 homogeneous, as well as self-dual, it will follow from the KV theorem that $\E(A)$ carries a formally real Jordan structure. 

Why should a model have either of these properties? It is instructive to look at the standard quantum model
associated with a finite-dimensional Hilbert space.  
As discussed above, $\V(A(\H)) \simeq \L_h(\H)$, the space of self-adjoint operators on $\H$. If $x$ is a unit vector in $\H$, let $p_x$ denote the corresponding rank-one orthogonal projection 
operator. Consider the trace inner product $\langle a, b \rangle = \Tr(ab)$ on $\L_h(\H)$:
by the spectral theorem, 
$\Tr(ab) \geq 0$ for all $b \in \L_{h}(\H)_+$ iff $\Tr(a p_x) = \langle a x, x \rangle\geq 0$ for all unit vectors $x$. So $\Tr(ab) \geq 0$ for all $b \in \E_+$ iff $a \in \L_{h}(\H)_+$, i.e., the trace inner product is self-dualizing. 
But this now leaves us with the question, {\em what does the trace inner product {\em represent}, probabilistically?}\\

\noindent{\bf The trace inner product as a bipartite state} Let $\bar{\H}$ be the conjugate Hilbert space to $\H$. Suppose $\H$ has dimension $n$. Any unit vector $\Psi$ 
in $\H \otimes \bar{\H}$ gives rise to a joint probability assignment to effects $a$ on $\H$ and $\bar{b}$ on $\bar{\H}$, namely $\langle (a \otimes \bar{b}) \Psi, \Psi \rangle$. Consider the 
maximally entangled {\em EPR state} for $\H \otimes \bar{\H}$ defined by the unit vector 
\[\Psi = \frac{1}{\sqrt{n}} \sum_{x \in E} x \otimes \bar{x} \in \H \otimes \bar{\H},\]
where $E$ is any orthonormal basis for $\H$. 
A straightforward computation shows that, for all $a, b \in \L_{h}(\H)$, 
$\langle (a \otimes \bar{b})\Psi, \Psi \rangle = \tfrac{1}{n} \Tr(ab)$.
In other words, {\em the normalized trace inner product {\em just is} the joint probability function determined by the pure state vector $\Psi$}! 
As a consequence, the state represented by $\Psi$ has a very strong correlational property: if $x, y$ are two orthogonal unit vectors with corresponding rank-one projections $p_x$ and $p_y$, we 
have $p_x p_y = 0$, so 
$\langle (p_x \otimes \bar{p_y}) \Psi, \Psi \rangle = 0$. 
On the other hand, 
$\langle (p_x \otimes \bar{p_x}) \Psi, \Psi \rangle = \tfrac{1}{n} \Tr(p_x) = \frac{1}{n}$. 
Hence, $\Psi$ {\em perfectly, and uniformly, correlates} every basic measurement (orthonormal basis) of $\H$ with its counterpart in $\bar{\H}$. Note that $\Psi$ is uniquely defined by this property. \\

\noindent{\bf Filters and homogeneity} {\red To say that the cone $\L_{h}(\H)_+$ is homogeneous means that any non-singular density operator can be obtained from 
any other by an order-automorphism. But in fact, something better is true: this order-automorphism can 
be chosen to represent a probabilistically reversible {\em physical} process, i.e., an invertible CP mapping. }
To see how this works, suppose $W$ is a positive operator on $\H$. Consider the pure CP mapping $\phi_{W} : \L_h(\H) \rightarrow \L_h(\H)$ given by 
$\phi_{W}(a) = W^{1/2} a W^{1/2}$.
Then $\phi_{W}(\1) = W$. If $W$ is nonsingular, so is $W^{1/2}$, so $\phi_{W}$ is invertible, with inverse 
$\phi_{W}^{-1} = \phi_{W^{-1}}$, again a pure CP mapping. Now given another nonsingular density operator $M$, 
we can get from $W$ to $M$ by applying $\phi_{M} \circ \phi_{W^{-1}}$.  

This is all well and good, but it leaves us with another question: {\em What does the mapping $\phi_{W}$ {\em represent}, physically?}
Suppose $W$ is a density operator, with spectral expansion $W = \sum_{x \in E} t_x p_x$. Here, $E$ is an orthonormal basis for $\H$ diagonalizing $W$, and $t_x$ is the eigenvalue corresponding to $x \in E$.
Then, for each vector $x \in E$,  
$\phi_{W}(p_x) = t_x p_x$, 
where $p_x$ is the projection operator associated with $x$.  Thus, if $M$ is a density operator corresponding to another state of the system, then the probability 
of $x$'s occurring in the (sub-normalized) state $\phi_{W}(M)$ is $t_x \Tr(Mp_x)$. 
In other words, $\phi_W$ acts as a {\em filter} on the test $E$: the {\em response} of each outcome $x \in E$ is attenuated by a factor $0 \leq t_x \leq 1$.

Thinking of the orthonormal basis $E$ as representing a set of alternative {\em channels} plus {\em detectors}, 
we can add a classical filter attenuating the response of one of the detectors --- say, $x$ --- by a fraction $t_x$. The discussion above tells us that we can achieve the same result by applying a suitable 
CP map to the system's state, {\em in advance} of the measurement. Moreover, this can be done independently for each outcome of $E$. 
This is illustrated below for 3-dimensional quantum system: $E = \{x,y,z\}$ is an orthonormal basis, representing three 
possible outcomes of a Stern-Gerlach-like experiment; the filter $\Phi$ acts on the system's state in such a 
way that the probability of outcome $x$ is attenuated by a factor of $t_x = 1/2$, while outcomes $y$ and $z$ are 
unaffected.  
\begin{center}
\begin{tikzpicture} 
\node[anchor=east] at (0,0) (source) {$M$};
\node[anchor=west] at (2,0) (splitter) {$\stackrel{\bigtriangledown}{\triangle}$}; 
\node[anchor=west] at (5,1) (detectorx) {$x$}; 
\node[red] at (7,1) {prob = $\frac{1}{2} \Tr(Mp_x)$};
\node[anchor=west] at (5,0) (detectory) {$y$};
\node at (7,0) {prob = $ \Tr(Mp_y)$};
\node[anchor=west] at (5,-1) (detectorz) {$z$};
\node at (7,-1) {prob = $\Tr(Mp_z)$};
\draw (source) edge[out=0,in=180,->] (splitter);
\draw (splitter) edge[out=0,in=180,->] (detectorx);
\draw (splitter) edge[out=0,in=180,->] (detectory);
\draw (splitter) edge[out=0,in=180,->] (detectorz);
\node at (1,0) (ring) {};
\draw[red] (1,0) circle(.2cm); 
\node[red] at (.5,-1) (filterlabel) {$\Phi$};
\draw[red] (filterlabel) edge[out=45,in=225,->] (ring);
\end{tikzpicture}\\
\vspace{.1in}
{Figure 1: $\Phi$ attenuates $x$'s sensitivity by $1/2$. }
\end{center}
\tempout{
\begin{center}
\begin{tikzpicture}
\draw[->] (0,0) -- (2,0) -- (3,0); 
\draw[->] (2,0) -- (2.2,1) -- (3,1);
\draw[->] (2,0) -- (2.2,-1) -- (3,-1); 
\draw (3.2,1) node{$x$};
\draw (3.2,0) node{$y$};  
\draw (3.2,-1) node{$z$};
\draw (2.7,1) node{$\Box$};
\draw (0,-.2) node{$\alpha$};
\end{tikzpicture} 
{Figure (a)}
\hspace{.2in}
\begin{tikzpicture}
\draw[->] (0,0) -- (2,0) -- (3,0); 
\draw[->] (2,0) -- (2.2,1) -- (3,1);
\draw[->] (2,0) -- (2.2,-1) -- (3,-1); 
\draw (3.2,1) node{$x$};
\draw (3.2,0) node{$y$};  
\draw (3.2,-1) node{$z$};
\draw (1.5,0) node{$\Box$};
\draw (1.5,.3) node{$\Phi_x$};
\draw (0,-.2) node{$\alpha$};
\end{tikzpicture}
{Figure (b)}
\end{center} 
}
If we apply the $\phi_{W}$ to the maximally mixed state $\tfrac{1}{n} \1$, we obtain $\tfrac{1}{n} W$. Thus, we can {\em prepare} $W$, up to normalization, by applying the process $\Phi_W$ to the maximally mixed state. 
What is more, as long as none of the eigenvalues $t_x$ is zero (that is, as long as $W$ is non-singular), $\Phi$ 
can be invertible, and hence, in the language of section 2, a p-reversible process
\\

\noindent{\bf Filters are Symmetric} Here is a final observation, linking these last two: The filter $\Phi_{W}$ is {\em symmetric} with respect to the uniformly correlating state $\Psi$, in the sense that \[\langle (\Phi_{W}(a) \otimes \bar{b}) \Psi, \Psi \rangle = \langle (a \otimes \bar{\Phi}_{W}(\bar{b}))\Psi, \Psi \rangle\]
for all effects $a, b \in \L_h(\H)_{+}$. 
As we'll soon see, this is basically {\em all that's needed} to recover the Jordan structure of finite-dimensional quantum theory: the existence of a conjugate system, with a uniformly 
correlating ``EPR"-like joint state, plus the possibility of preparing non-singular states by means of p-reversible filters that are symmetric with respect to this state.

\section{Conjugates and Filters} 

In order to abstract the features of QM discussed above, we first 
need to restrict our focus very slightly. Call a test space $(X,\M)$ {\em uniform} iff all tests $E \in \M$ have the 
same size, which we call the {\em rank} of $A$. The test spaces associated with Jordan (and so, in particular, with quantum) models all have this feature.\\

\noindent{\bf Definition:} Let $A$ be a uniform probabilistic model of rank $n$. A {\em conjugate} for $A$ is a model $\bar{A}$, plus a chosen isomorphism\footnote{By an isomorphism from a model $A$ to a model $B$, I mean a bijection $\phi : X(A) \rightarrow X(B)$ inducing a 
bijection $\M(A) \rightarrow \M(B)$, and such that $\alpha \in \Omega(A)$ iff $\alpha \circ \phi \in \Omega(B)$.}  $A \simeq \bar{A}$, taking $x \in X(A)$ to a $\bar{x} \in X(\bar{A})$, and a non-signaling bipartite state $\eta_A$ on $A$ and $\bar{A}$ such that $\eta_{A}(x,\bar{x}) = 1/n$ for all $x \in X(A)$. \\

If $E \in \M(A)$, we have $|E| = n$ and $\sum_{x,y \in E \times E} \eta_{A}(x,\bar{y}) = 1 = \sum_{x \in E} \eta(x, \bar{x})$.  Hence, $\eta_{A}(x,\bar{y}) = 0$ for $x,y \in E$ with $x \not = y$. Thus, $\eta_A$ establishes a 
perfect, uniform correlation between {\em any} test $E \in \M(A)$ and its counterpart, $\bar{E} := \{ \bar{x} | x \in E\}$, 
in $\M(\bar{A})$. Since $\eta$ is a bipartite state on $A$ and $\bar{A}$, it follows that its marginal, 
the uniformly or maximally mixed state $\rho(x) := \frac{1}{n}$, belongs to $\Omega(A)$. 
If $A = A(\H)$ is the quantum-mechanical model associated with an $n$-dimensional Hilbert space $\H$, then we can take $\bar{A} = A(\bar{\H})$ and define $\eta_{A}(x,\bar{y}) = |\langle \Psi, x \otimes \bar{y} \rangle|^2$, where $\Psi$ is the EPR state, as discussed in Section 3. 

So much for conjugates. We generalize the filters associated with pure CP mappings as follows:\\

\noindent{\bf Definition:} A  {\em filter} associated with a test $E \in \M(A)$ is a process  
$\Phi : \V(A) \rightarrow \V(A)$ such that for every 
outcome $x \in E$, there is some coefficient 
$t_x \in [0,1]$ with $\Phi(\alpha)(x) = t_x \alpha(x)$ for every state $\alpha \in \Omega(A)$.  \\

Equivalently, $\Phi$ is a filter iff the dual process $\Phi^{\ast} : \V^{\ast}(A) \rightarrow \V^{\ast}(A)$ satisfies 
$\Phi^{\ast}(\hat{x}) = t_x \hat{x}$ for each $x \in E$. Just as in the quantum-mechanical case, a filter independently attenuates the ``sensitivity" of the outcomes 
$x \in E$. We'll shortly see that the existence of a  conjugate, plus the {\em preparability} of arbitrary 
nonsingular states by {\em symmetric}, p-reversible filters, is enough to make $A$ a Jordan model. 
Most of the work is done by the easy Lemma 1, below. \\

\noindent{\bf Definition:} Let $\Delta = \{\delta_{x} | x \in X(A)\}$ be any family of states, indexed by outcomes $x \in X(A)$ with $\delta_{x}(x) = 1$.  I will say that a state $\alpha \in \Omega(A)$ is {\em spectral} with respect to 
$\Delta$ if there is a test $E \in \M(A)$ such that $\alpha = \sum_{x \in E} \alpha(x) \delta_x$. I'll call the model  
$A$ spectral with respect to $\Delta$ iff all its states are. \\

If $A$ is sharp and also spectral with respect to a set $\Delta$ of states, then 
$\Delta$ must be the unique set of states $\delta_x$ with $\delta_x(x) = 1$. Thus, for a sharp model, 
we can use the adjective ``spectral" without qualification. \\

\noindent{\bf Lemma 1:} {\em Let $A$ have a conjugate $(\bar{A},\eta_{A})$. Suppose that 
every non-singular state of $A$ is spectral with respect 
to the set of conditional states $\delta_{x} := \eta_{1|\bar{x}}$, $x \in X$. Then $A$ is sharp, and 
$\langle a, b \rangle := \eta_{A}(a,\bar{b})$
is a self-dualizing inner product on $\E(A)$, with respect to which 
$\V(A) \simeq \E(A)$. That is, $A$ is self-dual. }\\

\noindent{\em Proof:} That $\langle \, , \, \rangle$ is symmetric and bilinear follows from $\eta$'s being symmetric and non-signaling. Since $\hat{\eta}$ is a positive mapping,  $\hat{\eta}(\E(A)_+)$ is contained in $\V(A)_+$. By the spectrality assumption,  $\hat{\eta}(\E(A)_+)$ contains the interior of $\V(A)_+$. It follows (recalling that we are dealing with finite-dimensional spaces) 
that $\hat{\eta}(\E(A)_+) = \V(A)_{+}$, and, hence, that $\hat{\eta}$ is an order-isomorphism. The spectrality 
assumption now implies that 
every $a$ belonging to the interior of $\E(A)_+$ has a decomposition of the form $\sum_{x \in E} t_x x$ for 
some coefficients $t_x > 0$ and some test $E \in \M(A)$. It  follows that the such a decomposition is 
available (albeit with 
arbitrary coefficients) for any $a \in \E(A)$: for a sufficiently large value of $N$, $a + Nu$ belongs to 
the interior of $\E(A)_+$, and hence, has the desired decomposition, say 
$a + Nu = \sum_{x \in E} t_x x$ for some $E \in \M(A)$. Since 
$u = \sum_{x \in E} x$, we have $a = (a + Nu) - Nu = \sum_{x \in E} (t_x - N) x$. 

Now suppose $a \in \E$ with $a = \sum_{x \in E} t_x x$ for 
some $E \in \M(A)$ and coefficients $t_x \in \R$. Then 
\[
\langle a, a \rangle 
 =  \sum_{x, y \in E \times E} t_x t_y \eta_{A}(x, \bar{y}) 
 =  \frac{1}{n} \sum_{x \in E} {t_{x}}^2 \geq 0.
\]
This is zero only for $a = 0$, so 
$\langle \ , \ \rangle$ is an inner product. 
That this is self-dualizing, and makes $\E(A) \simeq \V(A)$, is straightforward (use the 
spectral decomposition plus the fact that $\hat{\eta}$ is an order-isomorphism). $\Box$ \\

If $A$ is sharp and has a conjugate $\bar{A}$, then the conditional state $ \eta_{1|\bar{x}}$ is the unique state $\delta_x$ with $\delta_x(x) = 1$, so the spectrality assumption in Lemma 1 is fulfilled if we simply say that $A$ is spectral. Hence, {\em a sharp, spectral model with a conjugate is self-dual.}
For the simplest systems, this is already enough to secure the desired representation in terms of a formally real 
Jordan algebra. Call $A$ a {\em bit}  
iff it has rank $2$ (all tests have two outcomes), and if every state $\alpha \in \Omega(A)$ can be expressed 
as a mixture of two sharply distinguishable states; that is, $\alpha = t \delta_x + (1 - t)\delta_y$ for some $t \in [0,1]$ and states $\delta_x$ and $\delta_y$ 
with $\delta_x(x) = 1$ and $\delta_y(y) = 1$ for some test $\{x,y\}$.  If a bit is sharp, then it is already spectral;  so it follows from Lemma 1 that 
if a sharp bit $A$ has a conjugate, it is self-dual. It follows easily 
 that $\Omega(A)$ must then be a ball of some finite dimension $d$. 
If $d$ is $2, 3$ or $5$, we have a real, complex or quaternionic bit. For $d = 4$ or $d \geq 6$, we have a non-quantum spin factor. 

Suppose now that $A$ has arbitrary rank, and satisfies the hypotheses of Lemma 1. 
If $\V(A)$ and, 
hence, $\E(A)$ are homogeneous, then, by the Koecher-Vinberg Theorem, $\E(A)$ carries a canonical Jordan structure. In fact, we can say something a little stronger \cite{Wilce12}: \\

\noindent
{\bf Theorem 1:} {\em Let $A$ be spectral with respect to a conjugate system $\bar{A}$.  
If $\V(A)$ is homogeneous, then there exists a canonical Jordan product on $\E(A)$ with respect to which $u$ is the Jordan unit. Moreover with respect to this product $X(A)$ is exactly the set of primitive idempotents, and $\M(A)$ is exactly the set of Jordan frames.} \\

The homogeneity of $\V(A)$ can be understood as a {\em preparability assumption:} it says 
that every nonsingular state can be obtained, {\em up to normalization}, from the 
maximally mixed state by a p-reversible process. 
\tempout{That is, if $\alpha \in \Omega(A)$, there is some 
such process $\phi$ such that $\phi(\rho) = p \alpha$ where $0 < p \leq 1$. One can think of the coefficient 
$p$ as the probability that the process $\phi$ will yield a nonzero result (more dramatically: will not 
destroy the system). Thus, if we prepare an ensemble of identical copies of the system in the maximally mixed state $\rho$ and subject them all to the process $\phi$, the fraction that survive will be about $p$, and these 
will all be in state $\alpha$. }  In fact, under the hypotheses of Lemma 1, the homogeneity of $\V(A) \simeq \E(A)$ follows  
from the mere existence of p-reversible filters with arbitrary non-zero coefficients. For if $a$ 
is in the interior of $\E(A)_+$, then $a = \sum_{x \in E} t_x x$ for some $E \in \M(A)$, with $t_x > 0$ for all $x \in E$. 
If $\Phi$ is a p-reversible filter with $\Phi(x) = t_x x$ $\forall x \in E$, then 
$\Phi(u) = a$. \\

\noindent{\bf Two paths to spectrality}  Some form of spectral decomposition for states 
is occasionally taken as an axiom \cite{Gunson, BMU}. However, spectrality can be derived 
from more transparent assumptions. 
\footnote{A different path to spectrality 
is charted in a recent paper \cite{CS} by G. Chiribella and C. M. Scandolo.}\\

\noindent{\bf Definition:}  A non-signaling bipartite state $\omega$ on probabilistic models $A$ and $B$ is {\em correlating} 
iff it sets up a perfect correlation between {\em some} test $E \in \M(A)$ and 
{\em some} test $F \in \M(B)$, 
in the sense that there exists 
a partial bijection $f : E \rightarrow F$ 
such that for all $x \in E, y \in F$, $\omega(x,y) > 0$ iff $f(x)$ is defined and $y = f(x)$.  \\

This is equivalent to saying that $\omega(x,f(x)) = \omega_1(x) = \omega_2(f(x))$, or 
$\omega_{2|x}(f(x)) = 1$, for $\omega_1(x) \not = 0$, for all $x \in E$. (Notice, too, that since $\omega$ must sum to $1$ over $E \times F$, $f$ must be non-empty.)  Using these observations and the law of total probability, 
we have \\

\noindent{\bf Lemma 2:} {\em Suppose $A$ is sharp. Any state arising as the marginal 
of a correlating bipartite state between $A$ and some model $B$, is spectral.} \\

 Let us say that $A$ satisfies the {\em correlation principle} iff every state of $A$ arises as the marginal of --- {\em dilates to} --- a correlating bipartite state.  We can paraphrase Lemma 2 as saying that if $A$ is sharp and satisfies 
this principle, then it is also spectral. 

 The correlation principle has an affinity with the purification postulate of \cite{CDP}, which requires that every state dilate to a {\em pure} state 
on a composite system. It is also related to the idea that, for every state $\alpha$, there should exist a 
{\em non-disturbing, recordable 
measurement}:  a test  $E \in \M(A)$ that can be made, without affecting $\alpha$, in such a way that 
the outcome can be recorded in the state of some ancillary system $B$. Prior to the test $E$, $A$ is in state $\alpha$ and $B$ is in some ``ready" state. After the test, the combined system is in
some joint, non-signaling state $\omega$. If the test is non-disturbing of $\alpha$, we must have 
$\omega_1 = \alpha$. If the outcome of $E$ was $x$, we suppose $B$ to be in a ``record state" $\beta_x$: if this record is accurate, we must have $\beta_x = \omega_{2|x}$, the conditional state of $B$ 
given $x$. If these record states are to be readable, there must exist a test $F$ on $B$ and a partial injection 
$f : E \rightarrow F$ such that $\beta_{x}(f(x)) = 1$ for every $x \in E$ with $\alpha(x) > 0$. Thus, 
$\omega$ correlates $E$ with $F$.  

Here is another, superficially quite different, way of arriving at spectrality. Let $A$ have a conjugate 
$(\bar{A},\eta_A)$. 
Call a transformation $\Phi : \V(A) \rightarrow \V(A)$ {\em symmetric} with respect to $\eta_A$ iff, 
for all $x, y \in X(A)$, 
$\eta_{A}(\Phi^{\ast} x, \bar{y} ) = \eta_{A}(x, \bar{\Phi}^{\ast} y)$.
Now let $\alpha = \Phi(\rho)$ where $\Phi$ is a symmetric filter on a test $E \in \M(A)$, say $\Phi(x) = t_x x$ for all $x \in E$. A direct computation then shows that $\alpha = \sum_{x \in E} t_x \delta_x$ (where $\delta_x = \eta_{1|\bar{x}}$). Thus, if every
nonsingular state is preparable by a symmetric filter, the spectrality assumption of Lemma 1 holds, and $A$ is 
self-dual. If the preparing filter can always be taken to be p-reversible, as well as symmetric, then $\V(A)$ is 
homogeneous, and we have a Jordan model. 
 On the other hand, as noted above, in the presence of spectrality, it's enough to {\em have} arbitrary p-reversible filters, as these allow one to prepare the spectral decompositions of arbitrary non-singular states. Thus, conditions 
(a) and (b), below, both imply that $A$ is a Jordan model. Conversely, one can show that any Jordan model satisfies 
both (a) and (b), closing the loop \cite{Wilce12}: \\

\noindent{\bf Theorem 2:} {\em The following are equivalent: 
\begin{mlist} 
\item[(a)] $A$ has a conjugate, and every non-singular state can be prepared by a p-reversible symmetric filter;
\item[(b)] $A$ is sharp, has a conjugate and arbitrary p-reversible filters, and satisfies the correlation principle;
\item[(c)] $A$ is a Jordan model.\\
\end{mlist}}

\noindent{\em Remark:} With some work, one can show that the assumptions of Lemma 1 imply a spectral uniqueness 
theorem, and hence, a functional calculus, for $\E(A)$. Call an effect $e \in \E(A)$ {\em sharp} iff 
there exists a state $\alpha$ with $\alpha(e) = 1$.  It is easy to show that $e$ must then have the form $e = e(D) := \sum_{x \in D} \hat{x}$ where $D \subseteq E$ for some $E \in \M(A)$. Call sharp effects $e_1,...,e_n$ {\em jointly orthogonal} iff $e_i = e(D_i)$ where $D_1,...,D_n$ are pairwise disjoint subsets of a single test $E \in \M(A)$.
 Then every $a \in \E(A)$ has a unique 
representation $a = \sum_{i=0}^{n} t_i e_i$ where $t_o > t_1 > .... > t_n$ and $e_1,...,e_n$ 
are jointly orthogonal sharp effects. Thus, for any function $f : \{t_o,...,t_n\} \rightarrow \R$, 
one can define $f(a) := \sum_{i} f(t_i) e_i$. In particular, $a^2 = \sum_i t_{i}^2 e_i$. There is 
now only one candidate for a Jordan product on $\E(A)$, namely, $a \dot b  = (a + b)^2 - a^2 - b^2$. 
If $\V(A)$ is also homogeneous, the KV theorem implies that this {\em is} a Jordan product. In particular, 
it is bilinear. An interesting problem is whether one can prove this {\em without} invoking the Koecher-Vinberg Theorem.

\section{Jordan Composites and Jordan Theories} 

A probabilistic theory is best understood as a category of probabilistic models 
and processes. In order to handle composite systems, one would like this to be a symmetric monoidal category. 
However, one wants to place some minimal restriction on how the monoidal product interacts with the 
probabilistic structure:\\

\noindent{\bf Definition:} A (non-signaling) {\em composite} of probabilistic models $A$ and $B$ is a model $AB$, equipped 
with a mapping $\pi : X(A) \times X(B) \rightarrow \V(AB)^{\ast}$, whereby outcomes $x \in X(A)$ and $y \in X(B)$ 
can be combined into a single effect $\pi(x,y) =: xy \in \V(AB)^{\ast}_{+}$. Moreover, we require that 
(i) $\sum_{(x,y) \in E \times F} \pi(x,y) = u_{AB}$ for all $E \in \M(A)$, $F \in \M(B)$, and 
(ii) $\forall \omega \in \Omega(AB)$, $\omega \circ \pi$ is a (non-signaling) bipartite state on $A$ and $B$. \\

By a {\em monoidal probabilistic theory}, I mean a symmetric monoidal category $\Cat$  in which 
objects are probabilistic models, morphisms are processes, and the monoidal product is a non-signaling composite in 
the above sense. Moreover, I require the monoidal unit to be the (obvious) trivial model $1$ with $\V(1) = \R$.
By a {\em conjugate} for $A \in \Cat$, I mean a conjugate in the sense defined earlier, but with the 
added restriction that $\bar{A} \in \Cat$ and $\eta_A \in \Omega(A \bar{A})$. \\

\noindent{\bf Theorem 3:} {\em Let $\Cat$ be a locally tomographic monoidal probabilistic theory in which every model $A$ is sharp, spectral, 
and has a conjugate in $\Cat$. Suppose also that (i) $\bar{\bar{A}} = A$, with $\eta_{\bar{A}}(\bar{a},b) = \eta_{A}(a,\bar{b})$, and (ii) if $\phi \in \Cat(A,B)$, then $\bar{\phi} \in \Cat(\bar{A},\bar{B})$. Then $\Cat$ has a canonical dagger compact structure, in which $\bar{A}$ is the dual of $A$ and $\eta_{A}$ is the co-unit.}\\

The proof is straightforward. Lemma 1 gives us a self-dualizing inner product on each of the spaces $\E(A)$, 
and also sets up canonical isomorphisms $\E(A) \simeq \V(A)^{\ast} \simeq \V(A)$, whence, we have a 
canonical inner product on the latter.  We can therefore regard $\Cat$ as a subcategory of the category 
$\FdHilb_{\R}$ of real finite-dimensional Hilbert spaces and linear mappings.  Using conditions (i) and (ii), 
plus local tomography, one shows that $\Cat$ inherits the $\dagger$-compact structure from the latter.

This raises two questions. The first is whether the local tomography assumption can be dropped. 
In a forthcoming paper  \cite{CCEJA}
(see also \cite{BGraW}),
 Howard Barnum, Matthew Graydon and I have shown that one can construct a dagger-compact category embracing 
real, complex and quaternionic quantum systems {\em at the same time}, at the cost of modifying the 
composition rule for complex quantum systems to include an extra classical bit, i.e., a two-valued 
superselection rule. (This has the function of allowing time-reversal to be a physical operation in 
 complex QM, as it is in the real and quaternionic cases). Composites in this category are generally 
 not locally tomographic. On the other hand, morphisms are not processes, in the sense defined above, 
 bu rather, certain positive mappings on the enveloping complex matrix algebras associated with 
 these Jordan algebras. 

The second question, which at present I cannot answer either, is what sort of converse, if any, holds for 
Theorem 3. That is, given a dagger-compact monoidal probabilistic theory consisting of finite-dimensional (uniform) probabilistic models, must these models satisfy the hypotheses of Lemma 1? Must they in fact be Jordan models?\\

\noindent{\bf Acknowledgments} 
Parts of this paper are based on talks given in workshops and seminars in Amsterdam and Oxford in 2014 and 2015, 
respectively. I wish to thank Drs. Sonja Smets and Bob Coecke for hospitality on these occasions. Work on this 
paper has been supported by a grant (FQXi-RFP3-1348) from the FQXi foundation.

\end{document}